\documentstyle[12pt]{article}

\def\be{\begin{equation}}
\def\ee{\end{equation}}
\def\bdi{\begin{displaymath}}
\def\edi{\end{displaymath}}
\def\br{\begin{eqnarray}}
\def\er{\end{eqnarray}}

\def\u2{\mid u\mid^2}

\def\ra{\rightarrow}

\def\RR{{\rm I\kern-.1567em R}}                              
 \def\CC{{\rm C\kern-4.7pt                                    
 \vrule height 7.7pt width 0.4pt depth -0.5pt \phantom {.}}} 
 \def\ZZ{{\sf Z\kern-4.5pt Z}}                                

\begin{document}

\begin{titlepage}
\vspace*{-2 cm}
\noindent

\vskip 3cm
\begin{center}
{\Large\bf Generalized integrability conditions and target space geometry }
\vglue 1  true cm
C. Adam$^1$ and  
J. S\'anchez-Guill\'en$^2$
\vspace{1 cm}

\vspace{1 cm}
{\footnotesize Departamento de F\'\i sica de Part\'\i culas,\\
Facultad de F\'\i sica\\
Universidad de Santiago\\
and \\
Instituto Galego de Fisica de Altas Enerxias (IGFAE) \\
E-15782 Santiago de Compostela, Spain}

\medskip
\end{center}

\normalsize
\vskip 0.2cm

\begin{abstract}
In some higher dimensional nonlinear field theories integrable subsectors 
with infinitely many conservation laws have been identified by imposing
additional integrability conditions. 
Originally, the complex eikonal equation was chosen as 
integrability condition, but recently further generalizations
have been proposed. Here we show how these new
integrability conditions may be derived from the geometry of the target
space and, more precisely, from the Noether currents related to a certain class
of target space transformations. 

\end{abstract}

\vskip 3.5cm

{\footnotesize $^1$email: adam@radon.mat.univie.ac.at} 

{\footnotesize $^2$email: joaquin@fpaxp1.usc.es}

\end{titlepage}

\section{Introduction }
Nonlinear field theories which allow for static, soliton type solutions
are relevant in different branches of physics, ranging from elementary
particle theory to condensed matter physics. Specifically for $3+1$
dimensional space-time and a two-dimensional target space the presence
of knot-like solitons can be expected provided that i) the condition of
finite energy requires the field to approach a constant value at 
spatial infinity and ii) the target space has the topology of the 
two-sphere $S^2$. In general, it is notoriously difficult to obtain
exact soliton solutions for such higher dimensional nonlinear field 
theories. On the other hand, in two dimensional space-time the concept of
integrability is known to simplify the calculation of exact solutions
significantly. A method to generalize the concept of integrability to
higher dimensions was therefore developped in \cite{AFSG}.
It was applied to the study of the Skyrme model \cite{Sky1}, which has
target space $S^3$ and consists, in addition to the usual sigma model
type quadratic term, of an additional quartic term in the Lagrangian
in order to circumvent the scaling instability (Derrick's theorem) 
and allow for static soliton
solutions. Further, the higher dimensional integrability was applied
to several restrictions of the Skyrme model to target space $S^2$,  
 like the Baby Skyrme model (in $2+1$ dimensional
space-time), the Faddeev--Niemi 
model \cite{Fad,FN1}, the
Aratyn--Ferreira--Zimerman (AFZ) model \cite{AFZ1,AFZ2}
(which only contains the quartic term with the appropriate power to avoid 
Derrick's theorem), or the Nicole model \cite{Ni1} (which only contains
the quadratic term, again with the appropriate power to avoid Derrick's
theorem);
see also \cite{BF1}, \cite{ASG2}. As all these models, except for the Skyrme
model\footnote{Besides their particular applications, they contain the basic 
ideas of Skyrme of topological and scale stability in a simplified form, 
which facilitates their analysis.}, have a two-dimensional target space, their
field content may be described by a complex field 
$u({\rm r},t)$, which is the case which we want to study in this paper. 
Integrability in this context amounts to the construction
of a infinite number of conserved currents. Indeed, if a current 
$K^\mu (u ,\bar u, u_\nu ,\bar u_\nu )$ can be found such that it obeys
the following three conditions (we abreviate $u_\mu \equiv \partial_\mu u$;
further, the bar denotes complex conjugation)
\be \label{cond1}
{\rm Im} (\bar u_\mu K^\mu ) =0
\ee
\be \label{cond2}
u_\mu K^\mu =0
\ee
\be \label{cond3}
\partial_\mu K^\mu =0   
\ee
then there exist the infinitely many conserved currents
\be \label{noet-cu}
J^G_\mu = i ( G_u K_\mu  - G_{\bar u} \bar K_\mu ) 
\ee
where $G$ is an arbitrary real function of $u$ and $\bar u$, and
$G_u \equiv \partial_u G$.   

It turns out that for the AFZ model a current $K^\mu$ can be determined
which obeys all three conditions (\ref{cond1}) - (\ref{cond3}) 
without further constraints. As a
consequence, the infinitely many currents (\ref{noet-cu}) are conserved
and generate infinitely many symmetries (the area-preserving diffeomorphisms
on the target space two-sphere $S^2$) 
of the AFZ model. In this model, therefore, integrability 
is realized and, further, infinitely many soliton solutions can be found
analytically
with the help of a separation of variables ansatz in toroidal coordinates,
which is indicated by the base space symmetries of the model \cite{AFZ2,BF1}. 

On the other hand, for the Baby Skyrme, Nicole, and Faddeev--Niemi models 
no current $K^\mu$ can be constructed which obeys all three conditions
(\ref{cond1}) - (\ref{cond3}) without further constraining the allowed 
fields $u$. It is, however, possible to find a current $K^\mu$ which obeys
conditions (\ref{cond1}) - (\ref{cond3}) provided that $u$ obeys the
additional constraint
\be \label{eik-eq}
\partial_\mu u \partial^\mu u =0,
\ee
the so-called complex eikonal equation. In the case of the Baby Skyrme model,
this condition just corresponds to the Cauchy--Riemann equations which
provide all the known instanton solutions of the two-dimensional sigma model
and, at the same time, all soliton solutions of the Baby Skyrme model.
More generally, the complex eikonal equation 
therefore defines integrable submodels for these three models where the
infinitely many currents (\ref{noet-cu}) are conserved. These infinitely
many conserved quantities are, however, no longer related to symmetries
of the submodels, because the eikonal equation is not of the Euler--Lagrange
type \cite{ASG2}.

Recently, another class of nonlinear field theories with integrable
submodels has been suggested by Wereszczy\'nski \cite{Wer1}, where a different,
``generalized''
first order constraint is imposed instead of the complex eikonal equation.
The construction of the constraint consists essentially in choosing
a vector-like quantity $\tilde K_\mu$, which
is a function of $u_\mu$ and $\bar u_\mu$ (but not on higher 
than first derivatives), and in imposing 
the constraint 
\be
u_\mu \tilde K^\mu =0.
\ee
For $\tilde K_\mu = u_\mu$
this leads to the eikonal equation, whereas for other choices a new 
integrability condition results.\footnote{Observe that these $\tilde K^\mu$
are, in general, different from the current $K^\mu$ of Eqs. (\ref{cond1}) -
(\ref{cond3}). In particular, they need no obey $\partial_\mu
\tilde K^\mu =0$ on-shell.} 
Further, some explicit 
Lagrangians were constructed in the same paper
with the help of the quantities $\tilde K_\mu$, 
and explicit soliton solutions were provided for
some particular members of this class of theories. These results
have the special interest of being one of the 
rare cases where explicit solutions for the integrable submodels have been
found. 

In our paper we want to provide a unifying view on all these models with
infinitely many conservation laws, both in the unconstrained case
and in the cases of the eikonal and the generalized constraints. Our approach
is based on a Noether current for target space transformations, essentially
the current (\ref{noet-cu}), and on the corresponding target space geometry.  
It turns out that the generalized integrability conditions and the 
corresponding explicit soliton 
solutions may be derived from a purely Lagrangian approach. Further, the
generalized integrability conditions turn out to
differ slightly from the eikonal
equation in that the former depend on the specific Lagrangian chosen,
whereas the eikonal equation only depends on the field contents (i.e., on
the complex field $u$). 
In Section 2 we present our approach and clarify its geometric significance.
In Section 3 we show that the soliton models of Wereszczy\'nski are covered
by our approach and easily rederive his results.  

\section{Conserved currents}
As said, we consider field theories where the field
content can be described by one
complex field $u$ and its complex conjugate $\bar u$. Concretely, we
allow for the class of Lagrangian densities
\be \label{g-lan}
{\cal L} (u ,\bar u ,u_\mu ,\bar u_\mu ) = F(a,b,c)   
\ee
where
\be
a=u\bar u \, ,\quad b=u_\mu \bar u^\mu \, ,\quad c= (u_\mu \bar u^\mu )^2
- u_\mu^2 \bar u_\nu^2 
\ee
and $F$ is at this moment an arbitrary real function of its arguments.
That is to say, we allow for Lagrangian densities which depend on the fields
and on their first derivatives, are Lorentz invariant, real, and obey the phase
symmetry $u\ra e^{i\phi} u$ for a constant $\phi \in \RR$. We could relax 
the last condition and allow for real Lagrangian densities which depend on 
$u$ and $\bar u$ independently, but this would just complicate the subsequent
discussion without adding anything substantial. Further, all models we want
to cover fit into the general framework provided by the 
class of Lagrangian densities 
(\ref{g-lan}), therefore we restrict our discussion to this class. 

For the currents $K^\mu$ we choose
\be \label{k-mu}
K^\mu = f(a) \bar \Pi^\mu
\ee
where $f$ is a real function of its argument, and
$\Pi^\mu$ and $\bar \Pi^\mu$ are the conjugate four-momenta of
$u$ and $\bar u$, i.e.,
\be
\Pi_\mu \equiv {\cal L}_{u^\mu} = \bar u^\mu F_b + 2 (u^\lambda \bar
u_\lambda \bar u_\mu - \bar u_\lambda^2 u_\mu )F_c.
\ee
The current of Eq. (\ref{k-mu}) automatically obeys the reality condition
(\ref{cond1}) for real Lagrangian densities. For the other two conditions  
(\ref{cond2}) and (\ref{cond3}) we find
\be \label{cond2a}
u_\mu K^\mu = f u_\mu^2 F_b
\ee
and, with the help of the equations of motion
\be \label{eom}
\partial^\mu \Pi_\mu ={\cal L}_u = \bar u F_a
\ee
and its complex conjugate,
\be \label{cond3a}
\partial^\mu K_\mu = f\left( M' \bar u u_\mu^2 F_b + u[ M' (bF_b + 2 cF_c )
+ F_a ] \right) 
\ee
where
\be
M \equiv \ln f 
\ee
and the prime denotes the derivative with respect to $a$. Before studying the 
conditions which make the r.h.s. of Eqs. (\ref{cond2a}) and (\ref{cond3a}) 
vanish, it is instructive to study the resulting expression for
$\partial^\mu J^G_\mu$, because this will clarify the geometry behind our
class of Lagrangian densities and the currents $J^G_\mu$. 
We find after a simple calculation
\br \label{div-jg}
\partial^\mu J^G_\mu & = & if \left( [( M' \bar u G_u + G_{uu} ) u_\mu^2 
 - ( M' u G_{\bar u} + G_{\bar u\bar u}) \bar u_\mu^2 ] F_b  \right.
\nonumber \\
&& \left. + \, (uG_u - \bar u G_{\bar u}) [ M' (bF_b + 2 cF_c ) +F_a ] \right)
\er

Now we want to ask under which circumstances this divergence may vanish.
If no constraints are imposed neither on the Lagrangian (i.e., on $F$)
nor on the allowed class of fields $u$, then we find the two equations for
$G$,
\be \label{geq1}
uG_u - \bar u G_{\bar u}=0,
\ee
and
\be \label{geq2}
M_a \bar u G_u + G_{uu} =0 \quad \Rightarrow \quad \partial_u [f(u\bar u)
G_u]=0 
\ee
together with  its complex conjugate. Eq. (\ref{geq1}) implies that
$G(u,\bar u) =\tilde G(u\bar u )$, and then Eq. (\ref{geq2}) leads
to the general solution
\be
G_u = k\frac{\bar u}{f}
\ee
where $k$ is a real constant. The corresponding current $J^G_\mu$ is the
Noether current for the phase transformation
\be \label{ph-tr}
u \ra e^{i \phi}u \quad ,\quad \bar u \ra e^{-i\phi}\bar u   
\ee
which is a symmetry of the Lagrangian by construction.

Next we restrict the possible Lagrangians by imposing (remember 
$M \equiv \ln f$)
\be
M_a (bF_b + 2 cF_c ) + F_a =0.
\ee
This equation can be solved easily by the method of characteristics and has the
general solution
\be \label{sol-cha}
F(a,b,c)= \tilde F (\frac{b}{f},\frac{c}{f^2})
\ee
which has, in fact, a nice geometric interpretation.
The point is that the resulting Lagrangian is a sigma model type of Lagrangian
which can be expressed entirely in terms of the target space geometry.
Indeed, trading the complex $u$ field for two real target space
coordinates $\xi^\alpha$, $u\ra (\xi^1 ,\xi^2 )$, the expressions on which
$\tilde F$ may depend can be expressed as follows. The first term is
\be
\frac{b}{f} = \frac{u_\mu \bar u^\mu}{f} = g_{\alpha \beta} (\xi)
\partial^\mu \xi^\alpha \partial_\mu \xi^\beta
\ee
where $\alpha =1,2$ etc, and the target space metric $g_{\alpha \beta}$ is
diagonal for the coordinate choice $\xi^1 ={\rm Re}\, u$, $\xi^2 = {\rm Im}
\, u$, i.e., $g_{\alpha \beta} = f^{-1}\delta_{\alpha \beta}$ . For the
second term we get
\be
\frac{c}{f^2}= \tilde \epsilon_{\alpha \beta} \tilde \epsilon_{\gamma \delta}
\partial^\mu \xi^\alpha \partial_\mu \xi^\gamma \partial^\nu \xi^\beta
\partial_\nu \xi^\delta
\ee
where
\be
\tilde \epsilon_{\alpha \beta}={\rm det}\, (g_{\gamma \delta}) 
\epsilon_{\alpha \beta}  \quad , \quad {\rm det}\, (g_{\gamma \delta}) 
= f^{-1}
\ee
and $\epsilon_{\alpha \beta}$ is the usual antisymmetric symbol in two
dimensions.
Observe that the two terms are different in that the first one, $b/f$, depends
on the target space metric, whereas the second one only depends on the
determinant of the target space metric.  

For Lagrangians which are of the form (\ref{sol-cha}) the condition 
that the divergence (\ref{div-jg}) vanishes only leads to Eq. (\ref{geq2})
for $G$, that is, to
\be \label{geq2a}
G_u = \frac{\bar H(u)}{f} \quad ,\quad G_{\bar u}=\frac{H(u)}{f}
\ee
where $H(u)$ is an analytic function of $u$ only. From what we found 
about the target space geometry, it will not come as a surprise that 
the solutions of Eq. (\ref{geq2a}) provide just the isometries of the
corresponding target space metric $g_{\alpha \beta}$. Indeed, from the
reality of $G$ and from the equality of the mixed second derivatives
$G_{u\bar u}$ one easily derives the equation
\be \label{heq}
H_u - \bar H_{\bar u} = M' (\bar uH - u\bar H) .
\ee
This equation always has the solution $H= ku$, independently of $M$, which
just corresponds to the symmetry under the phase transformation 
(\ref{ph-tr}).     Further solutions depend on the explicit form of
$M$, that is, $f$. E.g., for $M' =0$ (i.e., for a Euclidean metric on
target space), we find for $H$ the general solution
\be
H= k_1  + ik_2  + k_3 u \qquad {\rm for} \qquad k_i \in \RR
\ee
which generates the Euclidean group in $\RR^2$ (the isometries of
the flat, Euclidean metric in $\RR^2$). For $f=(1+u\bar u)^2$, which 
leads to the metric on $S^2$, we find for $H$
\be
H=    k_1 \frac{i}{2} (1-u^2)  + k_2 \frac{1}{2} (1+u^2) 
+k_3 u \qquad {\rm for} \qquad k_i \in \RR
\ee
which generates the modular transformations (the isometries of the 
metric for the two-sphere $S^2$, when the latter is expressed in the
coordinate $u$ via stereographic projection). For more generic expressions
for $f$, Eq. (\ref{heq}) does not provide further solutions and, 
consequently, the isometries are exhausted by the phase symmetries 
(\ref{ph-tr}).

Finally, we may further restrict the possible Lagrangians or the allowed field
configurations to achieve that the current divergence (\ref{div-jg}) vanishes
without requiring further restrictions on $G$. One way of achieving this
is by assuming $F_b \equiv 0$ identically, i.e.,
\be \label{sol-cha2}
F(a,b,c)\equiv \tilde F (\frac{c}{f^2}) 
\ee
(this has already been pointed out in \cite{BF1}, using a slightly 
different approach). The AFZ model is precisely of this type.
In this case the Lagrangian only depends on the determinant of the target
space metric, and it easily follows that a general $G$, which is now no longer
restricted (except for the condition of being real),  
is related to the area-preserving diffeomorphisms, i.e., $(iH \partial_u
+ {\rm c.c.}) $
generates area-preserving diffeomorphisms on functions of $u$ and $\bar u$, 
where $H\equiv fG_{\bar u}$. For the case of the two-sphere as target 
space, area-preserving diffeomorphisms and their generators and Noether
currents are discussed in detail e.g. in \cite{BF1}, \cite{ASG2},
\cite{FR1}.

Alternatively, we may make Eq. (\ref{div-jg}) vanish by imposing restrictions 
on the allowed field configurations $u$. In this case the currents $J^G_\mu$
are still the Noether currents of area-preserving diffeomorphisms, but these
transformations are no longer symmetry transformations of the pertinent
Lagrangians in general. We may either require that $u$ obeys the complex
eikonal equation (\ref{eik-eq}), or we may require that $u$ obeys the
(in general nonlinear) first order 
PDE which follows from the condition $F_b =0$ in cases
when this condition does not hold identically (i.e., for Lagrangians which do 
depend on the term $b=u^\mu \bar u_\mu$). The first case provides the
integrability condition for the integrable submodels of the Faddeev--Niemi,
Nicole and Baby Skyrme model, as was discussed, e.g., in \cite{AFSG}, 
\cite{ASG2}. The second case provides the generalized integrability
conditions which were introduced by Wereszczy\'nski in \cite{Wer1}, as we
shall discuss in the next section. Observe that the first condition, the
complex eikonal equation, is model independent, whereas the second condition
$F_b =0$ depends on the model, i.e., on the Lagrangian.

\section{Soliton models of Wereszczy\'nski}

Let us now specify the Lagrangian to
\be \label{We-La}
{\cal L}= \left( \lambda_1 \frac{b^3 }{f^3} + \lambda_2 \frac{bc}{f^3}
\right)^\frac{1}{2} = f^{-\frac{3}{2}}  
\left( \lambda_1 b^3  + \lambda_2 bc
\right)^\frac{1}{2}
\ee
where as above $f=f(a)$ and $\lambda_1$, $\lambda_2$ are two real constants.
This Lagrangian is of the type (\ref{sol-cha}). Further, 
the noninteger power in the Lagrangian is chosen precisely such as to
render the energies of field configurations scale invariant, avoiding 
thereby Derricks theorem and allowing for static, solitonic solutions.
In addition, it is equal to
the Lagrangian studied by Wereszczy\'nski (see Eq. (30) of Ref.
\cite{Wer1}) when the identifications
\be
f= G^{-\frac{2}{3}} \, , \quad \lambda_1 = \alpha + \beta +1 \, ,\quad
\lambda_2 = -\beta -1 
\ee
are made. The condition $F_b =0$ leads to the condition
\be
3\lambda_1 b^2 + \lambda_2 c =0
\ee
or, more explicitly,
\be \label{We-co}
3\lambda_1 (u_\mu \bar u^\mu )^2 + \lambda_2 ( (u_\mu \bar u^\mu )^2 - 
u_\mu^2 \bar u_\nu^2 ) =0
\ee
which coincides with the integrability condition Eq. (35) of Ref.
\cite{Wer1}.

Further, once the integrability condition (\ref{We-co}) is imposed, the
equation of motion is equivalent to the condition 
\be \label{eom-2}
\partial_\mu K^\mu =0,
\ee 
where $K^\mu$ is defined as before, $K^\mu = f \bar \Pi^\mu$ with
\be
\bar \Pi_\mu \equiv {\cal L}_{\bar u^\mu} = \frac{1}{2} f^{-\frac{3}{2}} 
\left( \lambda_1 b^3 + \lambda_2 bc\right)^{-\frac{1}{2}} [(3\lambda_1 
+ 2\lambda_2 ) b^2 u_\mu + \lambda_2 c u_\mu - 2 \lambda_2 b u_\nu^2 \bar 
u_\mu ] .
\ee
The equation of motion (e.o.m.) (\ref{eom-2})
coincides with Eq. (36) of Ref. \cite{Wer1}. 

Having unravelled the geometric nature of these further generalizations of 
integrability, let us finally derive in this framework
the explicit soliton solutions
of Weres\-zczy\'nski of the integrable submodels, that is, simultaneous 
solutions 
of the generalized constraint (\ref{We-co}) and of the e.o.m. (\ref{eom-2}).
Notice that in this systematic approach one also derives 
the corresponding Lagrangians (i.e., specific choices for
the target space metric function $f$ such that the field configurations 
solving the constraint solve at the same time the e.o.m.). 
The starting point for the construction of the solutions is the observation
that both the integrability condition Eq. (\ref{We-co}) and the e.o.m. 
(divergence condition) Eq. (\ref{eom-2}) have, in the static 
case, the conformal transformations on the base space
$\RR^3$ as symmetries. The symmetry under scale transformations is
obvious for both equations, whereas the symmetry under the remaining 
conformal transformations can be checked without difficulty (the
general method of calculating the symmetries of PDEs of
the above types is explained, e.g., in 
\cite{BF1}, \cite{ASG2}, \cite{Olv}). As a
consequence, both equations are compatible with a separation of variables
ansatz in toroidal coordinates. That is to say, if we introduce toroidal
coordinates  $(\eta ,\xi ,\varphi)$ via
\br
x &=&  q^{-1} \sinh \eta \cos \varphi \;\;, \;\;
y =  q^{-1} \sinh \eta \sin \varphi   \nonumber \\
z &=&  q^{-1} \sin \xi \quad ;  \qquad  q = \cosh \eta - \cos \xi .
\label{tordefs}
\er
then the ansatz 
\be \label{u-mn}
u = \rho (\eta)\, e^{ im\varphi + in\xi} \quad , \quad m,n \in \ZZ
\ee
is compatible with both equations and leads, in both cases, to an ODE for
$\rho (\eta)$. A detailed explanation for this ansatz and its relation
to the conformal symmetry of the static equations is provided in
\cite{BF1}. Further, fields $u$ within this ansatz can be interpreted
as Hopf maps $S^3 \to S^2$, and soliton solutions within this ansatz
are therefore topological in nature (``Hopf solitons''). For details on
the pertinent geometry and topology we refer, e.g., to 
\cite{BW1}, \cite{ASG1}, \cite{Wer2}. 

Now we proceed in two steps, analogously to the calculation in \cite{Wer1}. 
Firstly, we insert this ansatz into the 
integrability condition Eq. (\ref{We-co}) and find one solution for each
value of $m$ and $n$. Then we {\em assume} that this solution is also a 
solution of the e.o.m (\ref{eom-2}), insert it into the e.o.m.,  
and determine the target space
metric function $f$ accordingly (at this point our presentation differs 
slightly from the one chosen in \cite{Wer1}, where the solution for $f$ was
given at the beginning). 
This determination of $f$ is always possible, because the solution
$\rho (\eta)$ of the constraint (\ref{We-co}) allows to express $\eta$ in terms
of $\rho$, i.e., to express the e.o.m. as an ODE in the independent variable
$\rho$ and in the dependent variable $f$ (remember that $f=f(a)$ and 
$a=\rho^2$; we use the same letter $f$ also for $f(\rho)$, which should not
cause any confusion).   

We need  the gradient in toroidal coordinates
 \be \label{grad-3}
\nabla = (\nabla \eta)\partial_\eta 
+(\nabla \xi )\partial_\xi +(\nabla \varphi)\partial_\varphi
= q(\hat e_\eta \partial_\eta + \hat e_\xi \partial_\xi +
\frac{1}{\sinh \eta} \hat
e_\varphi \partial_\varphi  )
\ee
where $(\hat e_\eta  ,\hat e_\xi ,\hat e_\varphi )$ form an orthonormal
frame in $\RR^3$.  Further we need the relations
\be
\nabla \cdot \hat e_\eta = -\sinh \eta + \frac{1-\cosh \eta \cos 
\xi}{\sinh \eta} \, ,\quad \nabla \cdot \hat e_\xi = -2 \sin \xi \, ,\quad
\nabla \cdot \hat e_\varphi =0 .
\ee
Inserting now the ansatz (\ref{u-mn}) into the constraint equation 
(\ref{We-co}) we find after a brief calculation the equation
\be \label{We-co2}
\left( \frac{L_\eta}{\Gamma} \right)^4 -2 (\lambda -1) \left( 
\frac{L_\eta}{\Gamma} \right)^2 +1 =0  
\ee
where
\be
L\equiv \ln \rho \, ,\quad \Gamma \equiv \left( n^2 + \frac{m^2}{\sinh^2 
\eta }\right)^\frac{1}{2}
\ee
and
\be \label{def-lam}
\lambda \equiv  -\frac{2\lambda_2}{3\lambda_1}.
\ee
Eq. (\ref{We-co2}) is an algebraic second order equation for the quantity
$(L_\eta / \Gamma )^2$ with the solution
\be \label{We-co3}
\left( \frac{L_\eta}{\Gamma} \right)^2 = A^2 (\lambda) \equiv \lambda -1 + 
\sqrt{ \lambda^2 - 2\lambda }
\ee
and the condition that $A^2$ must be real and positive leads to the
restriction
\be \label{co-lam}
\lambda \ge 2.
\ee
[Remark: this restriction of $\lambda$ also determines the signs of
$\lambda_1$ and $\lambda_2$, which must have opposite signs according to
(\ref{def-lam}). The point is that the expression within the square root in 
the Lagrangian (\ref{We-La}) must be positive which, together with the
constraint (\ref{co-lam}), implies $\lambda_1 <0$ and $\lambda_2 >0$.]

Taking now the square root of Eq. (\ref{We-co3}) and choosing either of the 
two signs $\pm A$ on the r.h.s. we may integrate the expression for 
$L_\eta$ with the result $\rho^{(\pm)}$, where $\rho^{(-)} = 1/ \rho^{(+)}$
and
\be \label{sol-sol}
\rho^{(+)} = k \sinh^{A|m|} \eta \, \frac{\left( |n| \cosh \eta + \sqrt{
m^2 + n^2 \sinh^2 \eta} \right)^{A|n|}}{\left( |m| \cosh \eta + 
\sqrt{ m^2 + n^2 \sinh^2 \eta }\right)^{A|m|}}
\ee
where $k$ is a constant. This result coincides precisely with Eq. (24) of Ref.
\cite{Wer1}. It is interesting to observe that, for $A=1$, these 
field configurations also solve the static complex eikonal equation,
see \cite{Ada1}.

Now we insert this solution into the e.o.m. (\ref{eom-2}) and determine $f$
such that Eq. (\ref{eom-2}) holds. 
We restrict to the solutions $\rho^{(+)}$ and
find, after some calculation, for the spatial part $\vec K$ of the
current $K^\mu$
\be
\vec K = \frac{3}{2} {\cal A} f^{-\frac{1}{2}} q^2 u \rho \Gamma \left(
\hat e_\eta \Gamma A {\cal B} + i\left( n \hat e_\xi + \hat e_\varphi
\frac{m}{\sinh \eta}\right) {\cal C} \right)
\ee
where frequent use has been made of the relation $L_\eta = A\Gamma$. Further,
\br
{\cal A} &=& \left( 6\lambda A^2 (A^2 +1) - (A^2 +1)^3 \right)^{-\frac{1}{2}}
\nonumber \\
{\cal B} &=& - A^4 + 2 (2\lambda -1) A^2 + 2\lambda -1  
\nonumber \\
{\cal C} &=& (2\lambda -1) A^4 + 2 (2\lambda -1) A^2 -1
\er
are some functions of the parameter $\lambda$. 
For the divergence $\nabla \cdot \vec K$ we find, after some more calculation
(remember $M\equiv \ln f$),
\br
\nabla \cdot \vec K &=& \frac{3}{2} {\cal A} q^3 f^{-\frac{1}{2}} u\rho
A{\cal B} \left( -\frac{1}{2} \rho M_\rho A\Gamma^3 +2A \Gamma^3 +
\frac{\cosh \eta}{\sinh \eta} \left( \Gamma^2 -\frac{2m^2 }{\sinh^2 \eta}
\right) \right) \nonumber \\
&& - \frac{3}{2} {\cal A} q^3 f^{-\frac{1}{2}} u\rho \Gamma^3 {\cal C} .
\er

Before solving the conservation
equation $\nabla \cdot \vec K =0$, we restrict the
integers $m$ and $n$ to the case $m=n$. This we do because we need the
function inverse to $\rho^{(+)}(\eta)$. This would be extremely
complicated for $m \not= n$,
whereas it leads to the simple expression
\be \label{inv-rho}
\sinh \eta = \rho^{\frac{1}{|m|A}}
\ee
for $m=n$. Further we have for $m=n$ that
\be
\Gamma = |m| \frac{\cosh \eta }{\sinh \eta}
\ee
and find, with the help of the easily verified
identity ${\cal C} = A^2 {\cal B}$, that the condition $\nabla \cdot
\vec K =0$ leads to the equation
\be
\frac{1}{2} \rho M_\rho = 1 + \frac{1}{|m|A} \,
\frac{\sinh^2 \eta -1}{\cosh^2 \eta}
\ee
or, using Eq. (\ref{inv-rho}), 
\be
M_\rho = \frac{2}{\rho} \left( 1-\frac{1}{|m|A} \right) + \frac{4}{|m|A\rho}
\, \frac{ \rho^{\frac{2}{|m|A}}}{1+ \rho^{\frac{2}{|m|A}}}
\ee
with the solution
\be \label{cons-sol}
f= \rho^{2-\frac{2}{|m|A}} \left( 1+ \rho^{\frac{2}{|m|A}} \right)^2
\ee
which coincides exactly with the solution Eq. (40) of Ref. \cite{Wer1}. 

To finish, let us briefly comment on the construction of more complicated
scale-invariant models of the same type, 
which is discussed in Section 4 of Ref. \cite{Wer1}.
There the possibility of constructing further models was pointed
out, and it was observed that the integrability conditions for these
models have the same solutions (\ref{sol-sol}) within the ansatz  
(\ref{u-mn}) and, therefore, in this sense do not give rise to new 
field configurations. Here we just want to comment that these models
certainly are covered by our approach and, further, that it can be
easily understood from our methods why the corresponding integrability
conditions lead to the same solutions (\ref{sol-sol}). In fact, the
condition of scale invariance dictates that the corresponding Lagrangian
densities have to be of the form
\be
{\cal L}= f^{-\frac{3}{2}}\left( \alpha_k b^k + \alpha_{k-2} b^{k-2} c
+ \ldots + \alpha_1 b c^\frac{k-1}{2} \right)^\frac{3}{2k} \qquad
 \ldots \qquad k \, \, {\rm odd}
\ee
or
\be
{\cal L}= f^{-\frac{3}{2}}\left( \alpha_k b^k + \alpha_{k-2} b^{k-2} c
+ \ldots + \alpha_0  c^\frac{k}{2} \right)^\frac{3}{2k} \qquad
 \ldots \qquad k \, \, {\rm even}
\ee 
and the case discussed more explicitly in Section 4 of Ref. \cite{Wer1}
corresponds to $k=4$. Further, the integrability condition ${\cal L}_b =0$
leads in all cases to the equation
\be
\tilde \lambda_1 b^2 + \tilde \lambda_2 c =0
\ee
(where the $\tilde \lambda_i$ are some functions of the parameters $\alpha_j$
in the Lagrangian),
as may be checked easily. This integrability condition is the same as Eq. 
(\ref{We-co}), and, therefore, has the same solutions (\ref{sol-sol}). 
The only change is that the dependence of the parameter $A$ on the
original parameters in the Lagrangian is, of course, different for
different Lagrangians.

In conclusion, we see that the specific generalizations of the complex
eikonal equation, proposed in Ref. \cite{Wer1} as integrability
conditions in the sense of \cite{AFSG}, can be in fact understood
in general geometric terms, which allows to obtain directly the explicit
results of \cite{Wer1}, and to understand also its limitations.  \\ \\ \\
{\large\bf Acknowledgement:} \\
This research was partly supported by MCyT(Spain) and FEDER
(FPA2002-01161), Incentivos from Xunta de Galicia and the EC network
"EUCLID". Further, CA acknowledges support from the 
Austrian START award project FWF-Y-137-TEC  
and from the  FWF project P161 05 NO 5 of N.J. Mauser.

\end{document}